\documentclass[usenatbib]{mnras}
\usepackage[T1]{fontenc}
\usepackage{ae,aecompl}

\usepackage{graphicx}	
\usepackage{amsmath}	
\usepackage{amssymb}	
\usepackage{array}
\usepackage{physics}

\usepackage{multirow}
\usepackage{multicol}
\usepackage{blindtext}
\newcolumntype{?}{!{\vrule width 1pt}}
\newcommand{\Msun}{\,{\rm M}$_{\odot}$\,}
\newcommand{\Mpch}{\,{\rm Mpc}\,\ifmmode h^{-1}\else $h^{-1}$\fi}
\newcommand{\kpch}{\,{\rm kpc}\,\ifmmode h^{-1}\else $h^{-1}$\fi}

\newcommand{\kms}{\,{\rm km}\ s$^{-1}$\,}

\title[Superclusters from velocity divergence fields]{Superclusters from velocity divergence fields}
\author[Pe\~naranda-Rivera et al.]{
\parbox[t]{\textwidth}{
    {J. D. Pe\~naranda-Rivera $^1$,} 
    {D. L. Paipa-Le\'on$^{1}$,}
    {S. D. Hern\'andez-Charpak$^{1,2}$,}\\
    {J. E. Forero-Romero $^{1}$\thanks{E-mail: je.forero@uniandes.edu.co}}
}
\\\\
$^{1}$ Departamento de F\'isica, Universidad de los Andes, Cra. 1
  No. 18A-10 Edificio Ip, CP 111711, Bogot\'a, Colombia \\
$^{2}$ Center for Neuroprosthetics and Brain Mind Institute, Swiss
  Federal Institute of Technology (EPFL), CH-1015, Lausanne,
  Switzerland\\  
}

\date{Accepted XXX. Received YYY; in original form ZZZ}

\pubyear{2020}

\begin{document}
\label{firstpage}
\pagerange{\pageref{firstpage}--\pageref{lastpage}}
\maketitle

\maketitle
\begin{abstract}
Superclusters are a convenient way to partition and characterize the large scale structure of the Universe.
In this Letter we explore the advantages of defining superclusters as watershed basins in the divergence velocity field.
We apply this definition on diverse datasets generated from linear theory and  N-body simulations, with different grid sizes, smoothing scales and types of tracers.
From this framework emerges a linear scaling relation between the average supercluster size and the autocorrelation length in the divergence field, a result that holds for one order of magnitude from 10 \Mpch up to 100 \Mpch.
These results suggest that the divergence-based definition provides a robust context to quantitatively compare results across different observational or computational frameworks. 
Through its connection with linear theory, it can also facilitate the exploration of how supercluster properties depend on cosmological parameters, 
paving the way to use superclusters as cosmological probes.
\end{abstract}




\section{Introduction}

Superclusters are the largest structures that can be discriminated on the large scale structure of the Universe. 
From the early studies \citep{1983ARA&A..21..373O} through the most recent efforts to define our home supercluster Laniakea \citep{2014Natur.513...71T} a great variety of superclusters definitions have been proposed and explored, mostly motivated by the improvements in observational techniques and advances in computational models. 

Supercluster definitions span a broad conceptual range that includes: the manual segmentation of peculiar velocity fields  \citep{2014Natur.513...71T}, matter overdensity thresholds \citep{2015A&A...575L..14C}, percolation properties \citep{Bagchi_2017}, thresholds on multi-scale density fields \citep{Einasto_2019,2020A&A...641A.172E} and streamlines in the peculiar velocity data \citep{Dupuy_2019}, among others.
In general, all these definitions are different even if they operate on the same inputs. 
This means that the cross-algorithm comparison of supercluster properties, even as simple as their sizes, has to be done with caution.

A common feature in most of these studies is that the supercluster properties depend on the physical scale used to describe the
underlying density or velocity fields.
For instance, \cite{Dupuy_2020} reported strong changes in the supercluster abundance as a function of the smoothing scale used to define the velocity field.
\cite{2011A&A...531A.149S} also performed controled numerical experiments to measure how the cosmic web features, such as filaments, clusteras and voids, depend on the density perturbations at different scales.

Furthermore, the complexity in the algorithm used to actually find the superclusters in simulated or observed data also has an impact on the intepretation of the results.
A large number of free parameters could produce a large variability on the detected superclusters and raise the question as to what are the optimal combination of parameters values to define the superclusters \citep{Dupuy_2020}.

In this Letter we present a conceptual framework that allows us to reach a quantitative understanding of how the supercluster properties depend on the smoothing scale and develop a supercluster finding algorithm with low complexity.
The supercluster definition is based on the velocity divergence field while the supercluster finding algorithm uses a watershed concept.

This Letter is structured as follows.
In Section \ref{sec:divergence} we present our characterization of the velocity divergence field.
Next, in Section \ref{sec:watershed}, we describe the supercluster finding algorithm.
In Section \ref{sec:fields} we describe how we generate divergence fields on a grid to test our definitions and supercluster finding algorithm.
In Section \ref{sec:results} we present our main results to finally conclude in Section \ref{sec:conclusion} with future applications of the framework presented here.

\section{autocorrelation length in the divergence field}
\label{sec:divergence}
We use the velocity divergence field, $\div \textbf{v}$, as the main quantity to study over cosmological scales.
From it we define a dimensionless divergence field, $\delta(\textbf{r})$, as follows:

\begin{equation}
    \delta (\textbf{r})\equiv -\frac{1}{H_0 f} \div \bf{v},
    \label{eq:div}
\end{equation}
where $H_0$ is the Hubble parameter and $f$ is the growth rate of structure.
This is a purely computational definition motivated by the linear theory of structure formation where the equality between the two sides of Equation (\ref{eq:div}) is expected to actually hold when $\delta(\textbf{r})$ represents the matter overdensity.

We compute the power spectrum of this scalar field and denote it as $P_{\delta}(k)$. 
From this power spectrum we calculate the autocorrelation function as
\begin{equation}
    \xi_{\delta\delta} (r) = \frac{1}{2\pi^2}\int_{0}^{\infty} k^2 P_{\delta}(k)\frac{\sin kr}{kr} dk.
\end{equation}

From the autocorrelation function we define the autocorrelation length, $R_{\delta\delta}$, as the value of $r$ at which $\xi_{\delta\delta}(r)$ drops by a factor of ten from the value it has for $r\rightarrow0$. 
In practice, we compute the autocorrelation function in the interval $1\Mpch \leq r \leq 100 \Mpch$ and $R_{\delta\delta}$ is defined by the value at which
\begin{equation}
\xi_{\delta\delta}(R_{\delta\delta})  =\frac{1}{10} \xi_{\delta\delta}(1 \Mpch).    
\label{eq:autocorrelation}
\end{equation}

We show in the following sections how by using this definition we obtain a linear relationship between the autocorrelation length, $R_{\delta\delta}$, and the average supercluster size.

\begin{table*}
\begin{tabular}{c c c c c}\hline
Field &  $v_{max}$ & $L_b$ & $R_s$ & $N_c$\\
type & (km s$^{-1}$) & (\Mpch) & (\Mpch) & \\\hline
Gaussian & - & $720$ & \{$2$, $6$, $10$, $14$, $20$\} & $360$\\
Gaussian & - & $300$ & \{$3$, $9$, $12$, $15$, $18$\} & $100$\\
Gaussian & - & $1000$ & \{$10$, $15$, $20$, $30$, $40$\} & $100$\\
N-body & $150$ & $720$ & \{$6$, $12$, $30$, $48$, $60$\} & $120$\\
N-body & $200$ & $720$ & \{$5$, $10$, $20$, $30$, $40$\} & $144$\\
N-body & $300$ & $720$ & \{$2$, $6$, $10$, $14$, $20$\} & $360$\\\hline
\end{tabular}
\caption{Summary of the $30$ different dimensionless divergence fields used in this Letter. 
The Field type describes whether the field comes from the Abacus Simulations (N-body) or it is generated from the power spectrum (Gaussian). $v_{max}$ only applies to the N-body fields and refers to the minimum value of a halo maximum circular velocity to include it in the peculiar velocity field interpolation.
$L_b$ is the box size of the cubic volume used to compute the divergence. $R_{s}$ lists the gaussian smoothing (in the case of N-body fields) and the wavelength $1/R_s$ used to truncate the power spectrum (in the case of Gaussian fields). $N_c$ is the number of grid voxels on a side of the interpolated density field.}
\label{table:values}
\end{table*}

\begin{figure}
    \centering
    \includegraphics[width=250pt]{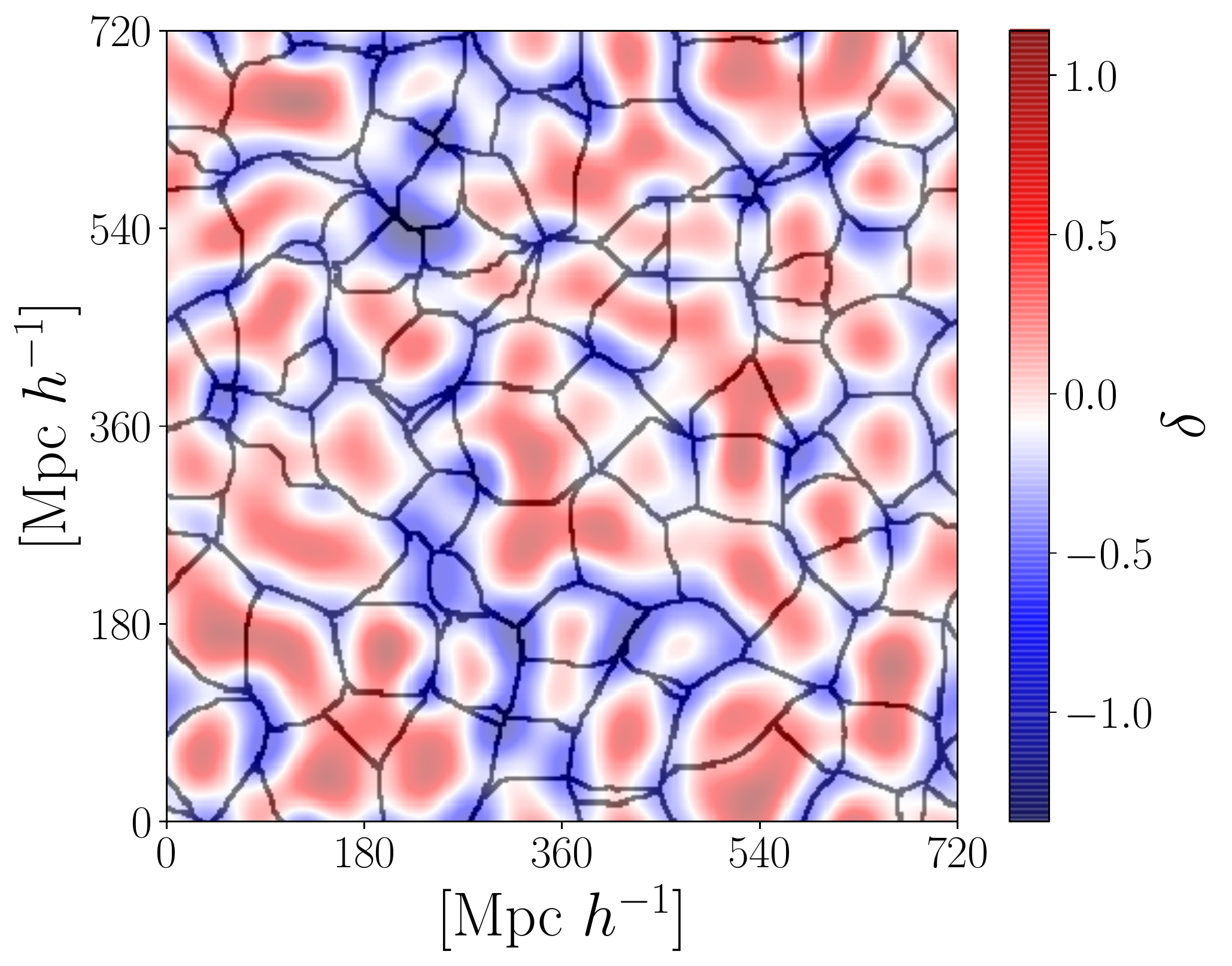}
    \caption{Slice across a 3D dimensionless divergence field (in colour) defined in Equation \ref{eq:div} together with the supercluster boundaries (in black) found by the watershed algorithm described in this Letter. 
    The divergence field is a Gaussian field with a truncation scale of $R_s=10$ \Mpch, sampled over $360^3$ voxels on a cubic box of $720$ \Mpch\ on a side.
    \label{fig:illustration}}
\end{figure}

\begin{figure*}
    \centering
    \includegraphics[width=240pt]{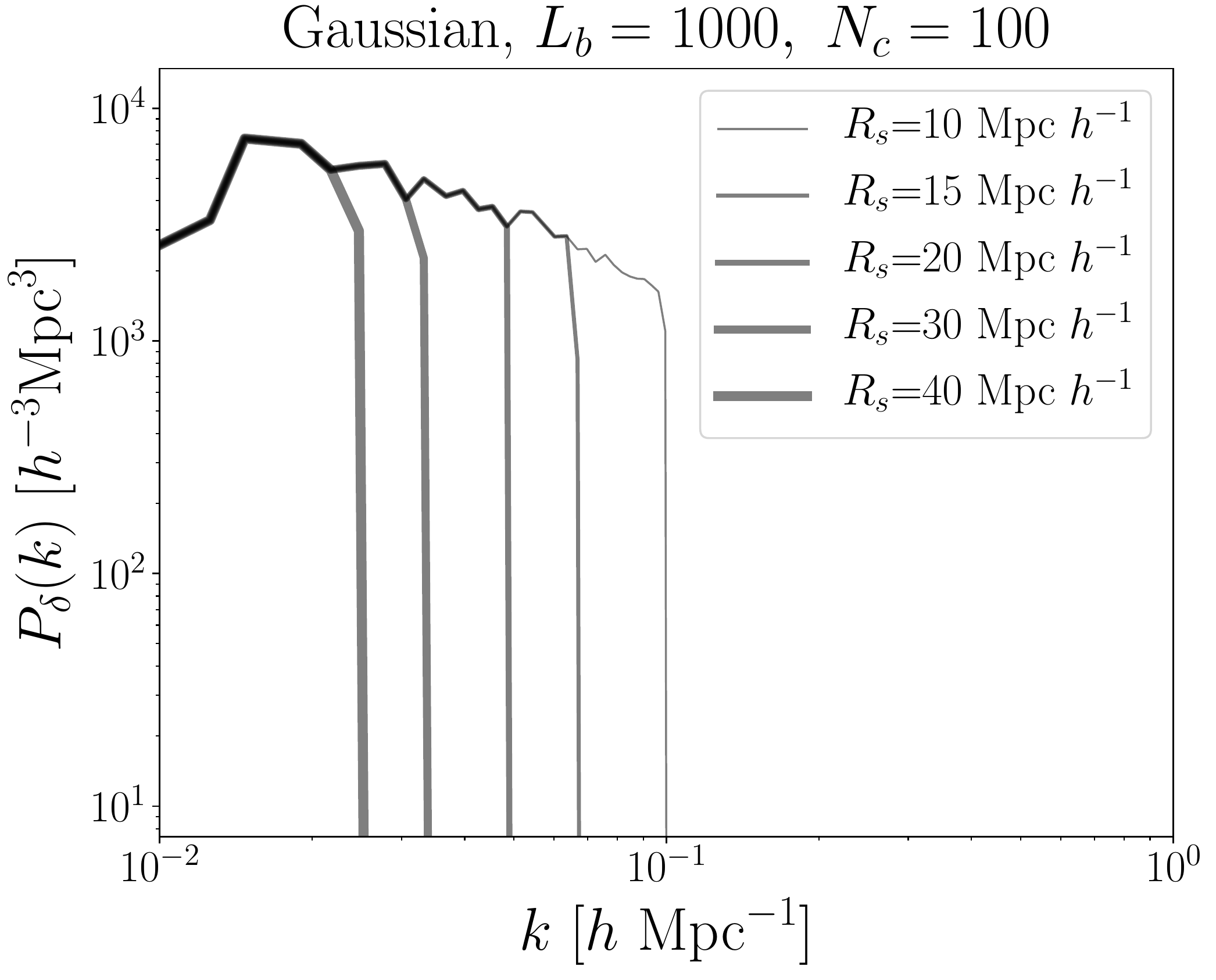}
    \includegraphics[width=240pt]{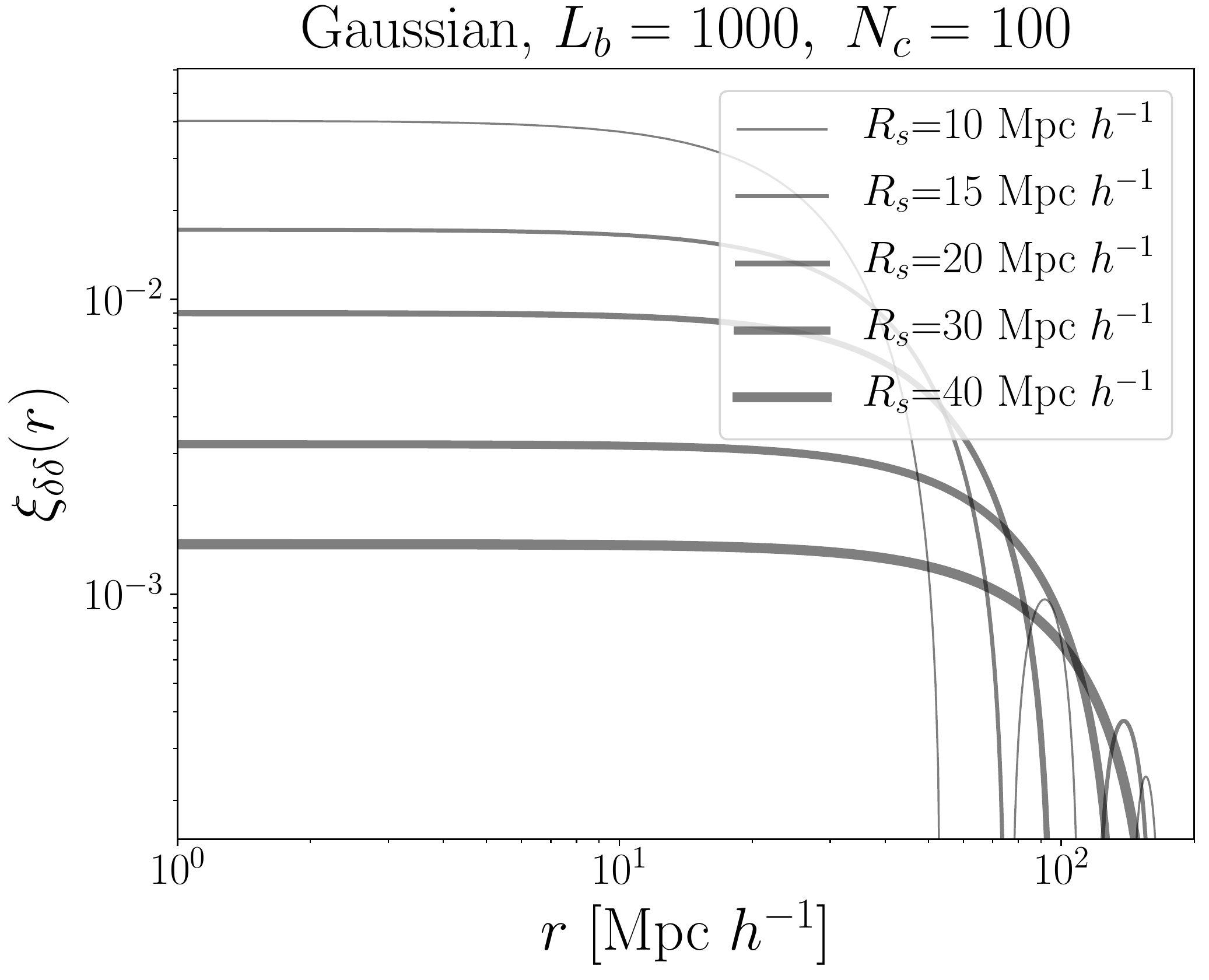}
    \caption{Power spectrum (left) and autocorrelation function (right) for dimensionless divergence fields computed from linearly extrapolated power spectrum
    The size of the cubic box is $L_b=1000$ \Mpch and the number of voxels is $100^3$.
    Each line corresponds to a different truncation wavelength in the power spectrum of $1/R_s$.
    \label{fig:gaussian}}
\end{figure*}

\begin{figure*}
    \centering
    \includegraphics[width=240pt]{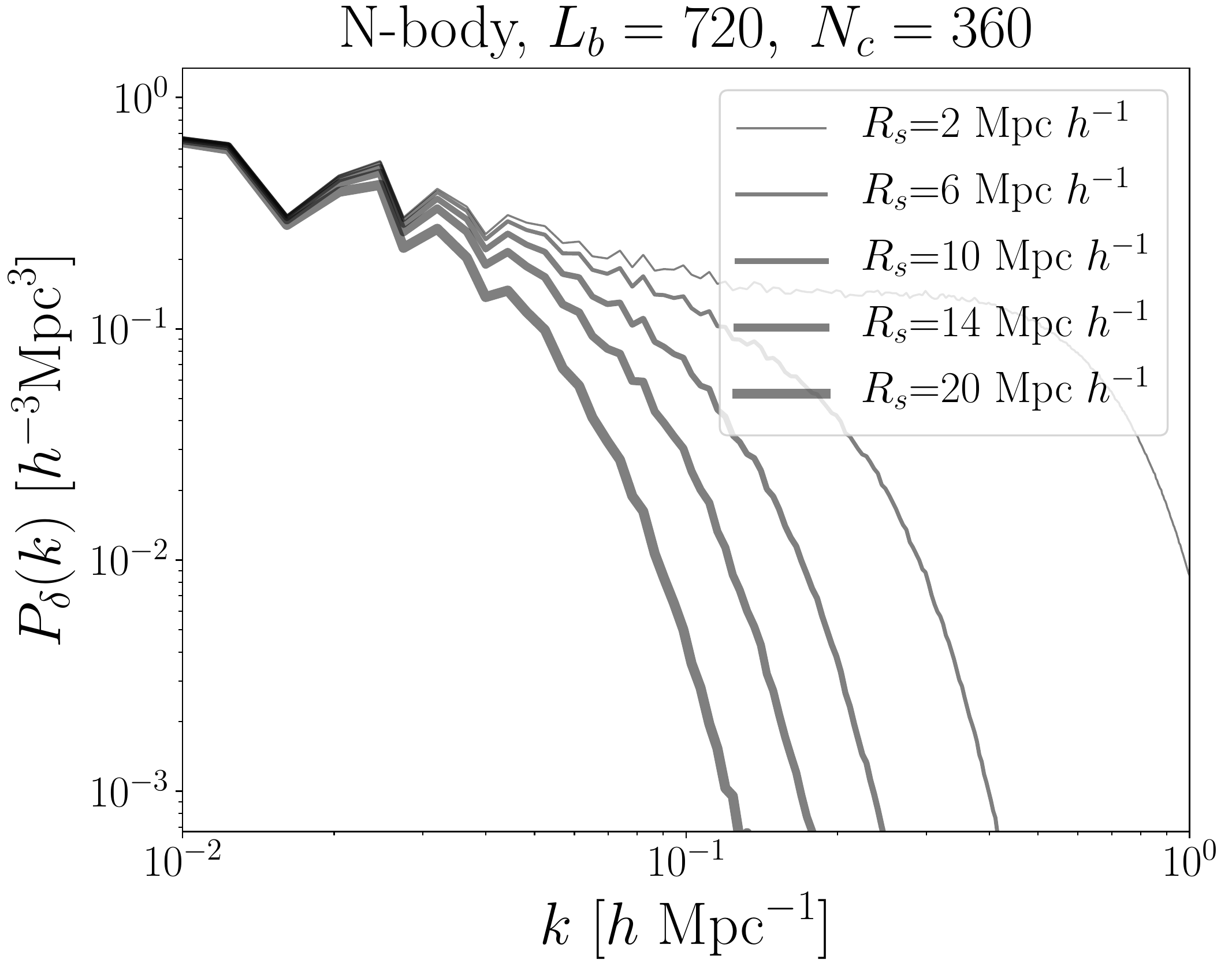}
    \includegraphics[width=240pt]{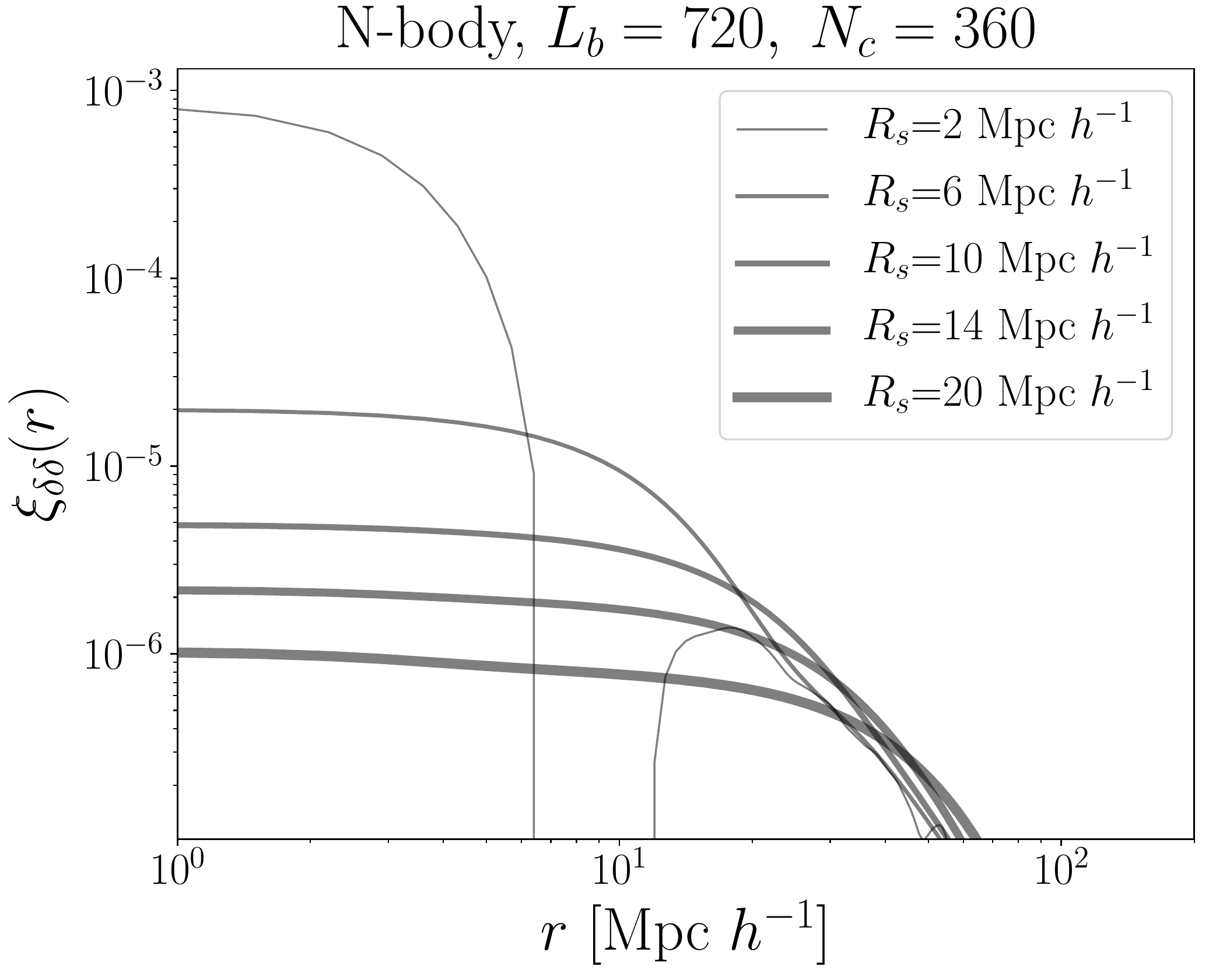} 
    \caption{Power spectrum (left) and autocorrelation function (right) for dimensionless divergence fields computed using velocity interpolation from N-body data.
    The size of the cubic box is $L_b=720$ \Mpch and the number of voxels is $360^3$.
    Only dark matter haloes with maximum circular velocity larger than $300$ \kms were included in the velocity interpolation.
    Each line corresponds to a different Gaussian smoothing with length scale $R_s$.}
    \label{fig:nbody}
\end{figure*}

\section{Watershed Superclusters}
\label{sec:watershed}

We find the superclusters using a watershed algorithm \citep{beucher1993morphological} on the dimensionless divergence fields.
The algorithm segments the whole volume by assigning each voxel to a unique supercluster. 

Our implementation works as follows. 
We sweep the voxels in the dimensionless divergence grid from the highest values to the lowest, i.e. from the high density regions with converging flows to the lowest density regions with divergent flows.
For the $i$-th voxel under consideration we check whether its 26 neighbors have already been assigned to a group. 
If all the neighbors are unassigned, then this $i$-th voxel starts a new group; if the majority of already assigned voxels belongs to the $n$-th group, then this voxel belongs to that group.
In case of a draw between groups, the voxel is assigned to the supercluster with the lowest $n$ value.
At the end of the sweep all voxels have been assigned to a group. 
In all our calculations we take periodical boundary conditions into account.  

Finally, each group found by the algorithm is interpreted as a supercluster 
with a volume, $V_s$, calculated by the total volume of the voxels belonging to it.
From the volume $V_s$ we compute the equivalent radius $R_{\rm eq}$, that is the radius of the sphere with the same volume, 
   $ R_{\rm eq} = \left(\frac{3}{4\pi}V_{s}\right)^{1/3}$.
We describe the supercluster population in a divergence field by the average value of the equivalent radius, $\langle R_{\rm eq}\rangle$.

Once the divergence field is provided, this algorithm  does not have free parameters based either on divergence, density or distance thresholds. 
The algorithm has a single free parameter: the number of neighbors to make the search for assigned superclusters.
In our case we use $26$, but our tests show that using lower values down to $6$ neighbors do not impact the main scaling results reported in this Letter. 
A lower number of neighbors only results in a larger tail of small superclusters composed by a handful of voxels.

Figure \ref{fig:illustration} illustrates the results of the watershed algorithm. 
This shows how the typical supercluster size follows the typical size of the features in the divergence field.

\section{Divergence Fields on a Grid}
\label{sec:fields}

We use dimensionless divergence fields computed on a cubic grid. 
We generate them in two different ways.
The first corresponds to Gaussian fields with a truncated power spectrum coming from linear theory.
The second kind are fields generated by interpolation of biased tracers and smoothed with a Gaussian filter, with the tracers coming from N-body cosmological simulations.
In both cases the cosmological parameters correspond to the
$\Lambda$CDM Planck 2015 cosmology \citep{2016A&A...594A..13P} with a Hubble parameter of $H_0=67.8$ km s$^{-1}$ Mpc$^{-1}$, a matter density parameter $\Omega_m=0.308$ and a spectral index $n_s=0.968$.
We use a total of $30$ different divergence fields, summarized in Table \ref{table:values}, generated as follows.

\subsection{Gaussian fields with truncated power Spectrum}

We start by generating a density field with a power spectrum that follows no-wiggle parameterization of \cite{1998ApJ...496..605E}.
The power spectrum is extrapolated at $z=0$ using linear theory and truncated at $k_{\rm max}=1/R_s$.
This procedure can be thought as erasing fluctuations smaller than $R_s$.
This density field is generated over a cubic grid with $N_c^3$ voxels over a cube of length $L_s$ on a side.
From this Gaussian density field we compute  the overdensity and directly use it as the dimensionless divergence field.
We generate $15$ different Gaussian fields by changing the values for $L_b$, $N_c$ and $R_s$. 
Table \ref{table:values} summarizes the different values for those parameters.

\subsection{N-body fields from biased tracers with Gaussian smoothing}

We use the public Friend-of-Friends catalogs at redshift $z=0.1$
from the Abacus Project simulations \citep{2018ApJS..236...43G} to interpolate the haloes' peculiar velocity over a grid with $N_s^3$ voxels.
The simulations were performed on a cube of side length $720$\ \Mpch with
$1440^3$ particles, corresponding to a DM particle mass resolution of $\sim 1 \times 10^{10}$ \Msun.
We start by selecting haloes with maximum circular velocity larger
than $v_{max}$.
Then we assign to each voxel the value of the average velocity of the haloes in that voxel, which is equivalent to a Nearest Grid Point interpolation scheme. 
If the voxel is empty we assign a value of $0$\kms.
After this interpolation step we smooth the velocity field using a Gaussian filter of physical scale $R_s$.
We generate $15$ different N-body based divergence fields by changing the values for $v_{max}$, $N_c$ and $R_s$. 
Table \ref{table:values} summarizes the different values for those parameters.

\begin{figure}
    \centering
    \includegraphics[width=220pt]{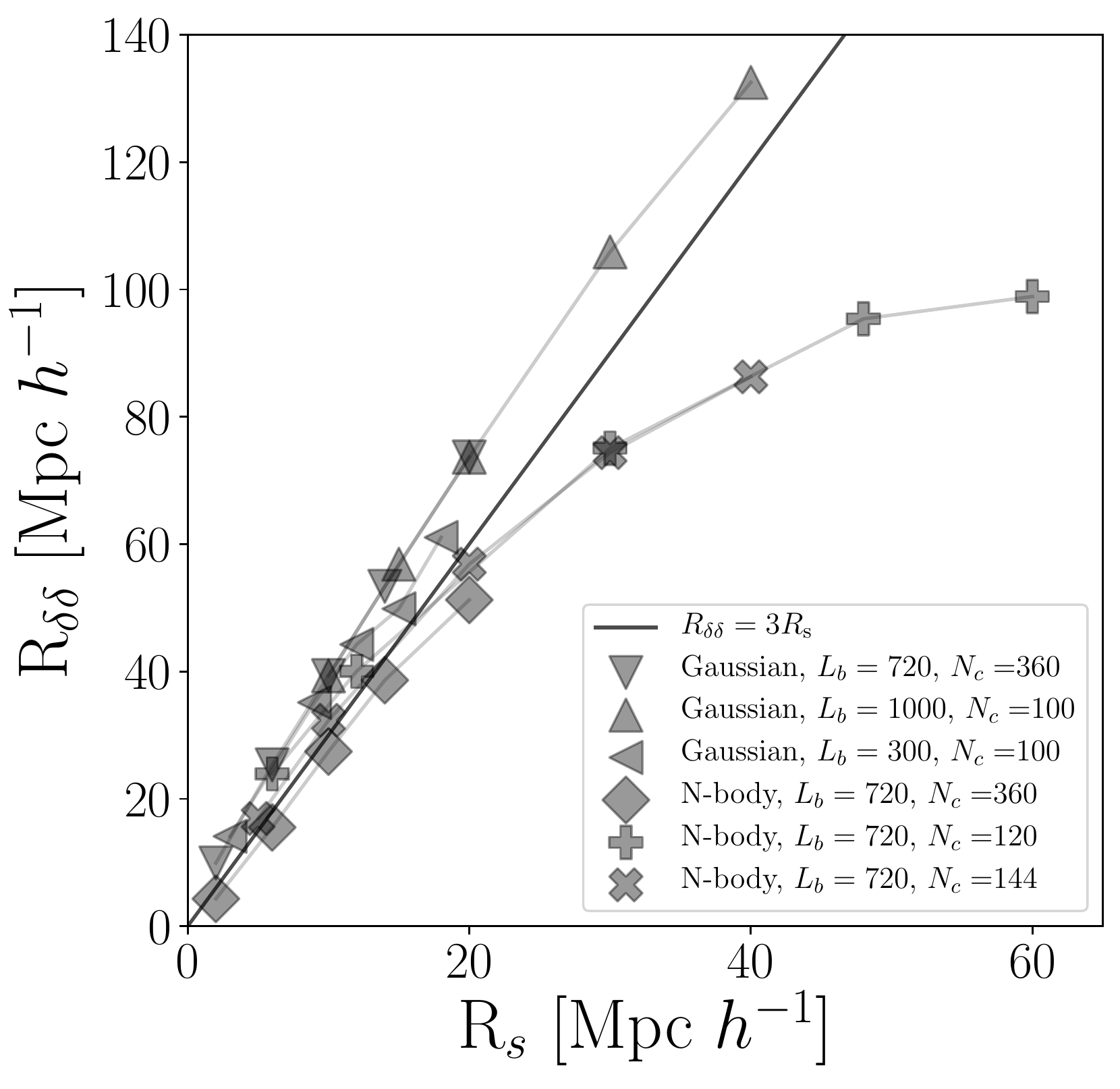}
    \caption{Correlation length $R_{\delta\delta}$ as a function of the truncation scale $R_s$.
    Both for Gaussian and N-body fields the relationship between the two scales is close to linear below $R_{s}=20$ \Mpch.}
    \label{fig:correlation length}
\end{figure}

\section{Results}
\label{sec:results}

We start by quantifying the effect of the truncation/smoothing scale.
Figure \ref{fig:gaussian} and Figure \ref{fig:nbody} show the power spectrum and the corresponding autocorrelation function.
In both cases, the right panel shows that the autocorrelation length increases with larger $R_{s}$ values.

Figure \ref{fig:correlation length} shows the relationship between the truncation/smoothing scale $R_s$ and the autocorrelation length $R_{\delta\delta}$.
We find that below $R_{s} < 20$\Mpch the two scales almost follow a linear proportionality.
Beyond that scale, the Gaussian fields continue that linearity up to $R_{\delta\delta}\approx 130$\Mpch, while the N-body fields saturates around $R_{\delta\delta}\approx 100$\Mpch.

\begin{figure}
    \centering
    \includegraphics[width=220pt]{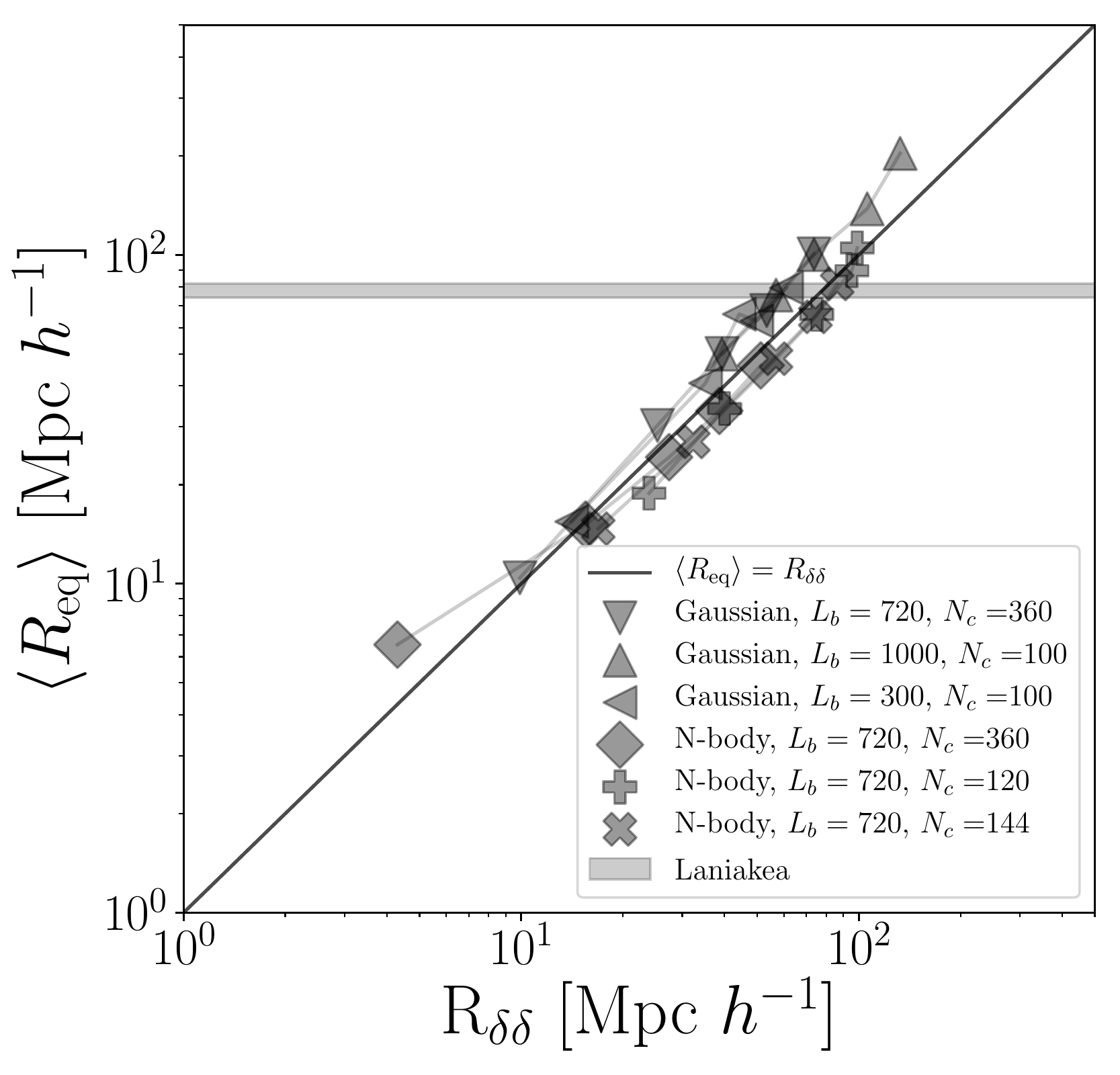}
    \caption{Average equivalent radius as a function of the autocorrelation length in their parent divergence field.
    The relationship between the two quantities is close to linear for autocorrelation lengths between $10$ \Mpch and $100$ \Mpch. \label{fig:money_plot}}
\end{figure}

\begin{figure}
    \centering
    \includegraphics[width=220pt]{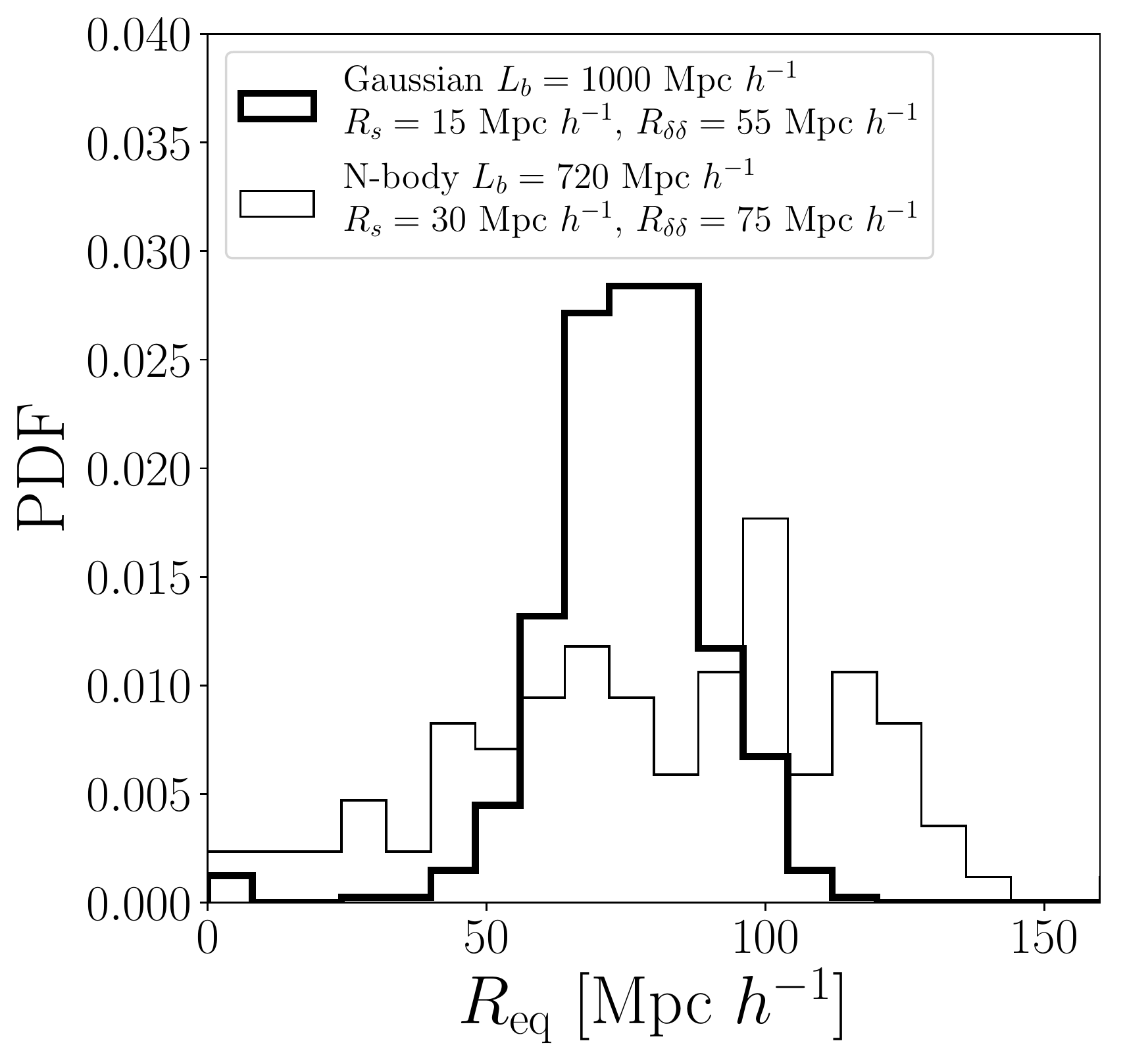}
    \caption{Equivalent radius distributions from dimensionless divergence fields derived from a Gaussian random field and a N-body simulation.
    In both cases the average value for the equivalent radius is close to $80$\Mpch. 
    The distributions shapes are different; the Gaussian field produces a narrower, closer to Gaussian, distribution than the N-body field. }  
    \label{fig:Nclusters}
\end{figure}

Our most important result is summarized in Figure \ref{fig:money_plot}.
The plot shows that 
the average supercluster size linearly follows the autocorrelation length as defined in Equation \ref{eq:autocorrelation}.
The linear correlations holds both for Gaussian and N-body fields over one order of magnitude from $10$\Mpch up to $100$\Mpch.
In other words, the autocorrelation formalism over the velocity divergence fields is able to estimate the typical supercluster size.
We parameterize this strong linear dependence  as 
$\langle R_{\rm{eq}}\rangle=\alpha R_{\delta\delta}$, the best fit gives $\alpha=1.38\pm0.03$ for Gaussian fields and $\alpha=0.93\pm 0.01$ for N-body fields.

This linear scaling relation with $\alpha\approx 1$ depends on the constant used in the definition in Equation \ref{eq:autocorrelation}). 
Using values larger than $1/10$ in that definition breaks the linear scaling above $R_{\delta\delta}\approx 40$\Mpch for N-body fields. 
Furthermore, the regions where the linear scaling still holds, show values of $\alpha$ smaller than unity. 
Using values smaller than $1/10$ also breaks the linearity. 
It also has the inconvenient of probing scales where the autocorrelation function might oscillate, making ambiguous the autocorrelation length definition.
    
We finalize this Section by showing how the supercluster size distribution differs in Gaussian fields compared to the N-body fields.
Figure \ref{fig:Nclusters} shows the probability density function for the equivalent radius distributions from two different divergence fields.
We pick these two distributions because they have similar mean values of $75$\Mpch\ and $81$ \Mpch\ for the Gaussian and N-body fields, respectively. 
However, the distribution in the N-body field is wider (standard deviation of $34$ \Mpch) than it is for the Gaussian field (standard deviation of $14$ \Mpch), meaning that there are more superclusters with extreme values, say above $100$\Mpch, in the non-linear divergence field than in the Gaussian field.

What do these results have to say about Laniakea?
The horizontal stripe in Figure \ref{fig:money_plot} shows recent estimates of Laniakea's equivalent radius \citep{Dupuy_2019}. 
If Laniakea is to be considered as an average supercluster, then our results predict that the autocorrelation length in the divergence fields from the Cosmicflows data should be on the order of 60 \Mpch\ to 85 \Mpch.
Having a measurement of the autocorrelation length on the divergence field from Cosmicflows-2, would allow to quantify to what extent Laniakea is a typical supercluster in a cosmological context.

\section{Conclusions}
\label{sec:conclusion}

In this Letter we presented the peculiar velocity divergence field as a central element to define and understand superclusters.
We computed its autocorrelation function to give a precise definition of an autocorrelation length. 
To define the superclusters we performed a watershed partition over the same velocity divergence field. 
We tested this picture using different velocity divergence fields (Gaussian fields and fields from N-body simulations) computed over different grid sizes, smoothing scales and types of dark matter halos as tracers.

We found a linear scaling relation between the average supercluster size and the autocorrelation length. 
This result holds for one order of magnitude from $10$\Mpch up to $100$\Mpch.
Comparing the full size distribution in the Gaussian and N-body divergence fields, selected to have average supercluster size similar to Laniakea, we found that the distributions are narrower and more symmetric in the Gaussian field.

The concepts we have presented here are straightforward to apply both to observational data and simulations.
For instance the velocity divergence field has already been used in CosmicFlows-2 reconstruction data to define cosmic web elements \citep{2015MNRAS.452.1052L} and could thus be used to find superclusters and compute its autocorrelation length.

There are multiple directions that could be addressed in future work. 
For instance, measuring to what extent the linearity between scales  in Figure \ref{fig:correlation length} depends on the box size, comparing supercluster sizes from different definitions that take similar data as an input \citep{2020A&A...641A.172E}, finding scaling relationships between sizes, 3D shapes, maximum divergence depth, and the dependence of supercluster properties on cosmological parameters together with its redshift evolution.

Finally, studying the link between divergence based superclusters, their density field \citep{2020A&A...641A.172E}
and graph properties \citep{2020MNRAS.498L.145G}
 computed on top of the galaxy spatial distribution could provide a potential way to define superclusters where velocity information is not available and also quantifying the relationship between large scale structure topology and dynamics.

\section*{Data Availability Statement}
The N-body data used to build the interpolated divergence fields are available through \url{https://lgarrison.github.io/AbacusCosmos/}. 
The code to generate the gaussian random fields and find the superclusters is available at \url{https://github.com/astroandes/WatershedDivergenceSuperclusters}

\bibliographystyle{mnras}

\end{document}